\documentclass[]{revtex4-1}
\usepackage{graphicx}
\usepackage{amssymb}
\usepackage{color}
\usepackage{amsfonts}
\usepackage{epstopdf}

\newcommand{\aver}[1]{\langle #1 \rangle}

\begin{document}


\title{Performance losses of drag-reducing spanwise forcing at moderate values of the Reynolds number}
\date{\today}

\author{Davide Gatti}
\affiliation{Center of Smart Interfaces, Technische Universit\"at Darmstadt, Alarich-Wei\ss -Stra\ss e 10, 64287 Darmstadt, Germany}
\email{gatti@csi.tu-darmstadt.de}
\author{Maurizio Quadrio}
\affiliation{Dept. of Aerospace Sciences and Technologies, Politecnico di Milano, Via La Masa 34, 20156 Milano, Italy}
\email{maurizio.quadrio@polimi.it}

\begin{abstract}
A fundamental problem in the field of turbulent skin-friction drag reduction is to determine the performance of the available control techniques at high values of the Reynolds number $Re$. We consider active, predetermined strategies based on spanwise forcing (oscillating wall and streamwise-traveling waves applied to a plane channel flow), and explore via Direct Numerical Simulations (DNS) up to $Re_\tau=2100$ the rate at which their performance deteriorates as $Re$ is increased. To be able to carry out a comprehensive parameter study, we limit the computational cost of the simulations by adjusting the size of the computational domain in the homogeneous directions, compromising between faster computations and the increased need of time-averaging the fluctuating space-mean wall shear-stress.

Our results, corroborated by a few full-scale DNS, suggest a scenario where drag reduction degrades with $Re$ at a rate that varies according to the parameters of the wall forcing. In agreement with already available information, keeping them at their low-$Re$ optimal value produces a relatively quick decrease of drag reduction. However, at higher $Re$ the optimal parameters shift towards other regions of the parameter space, and these regions turn out to be much less sensitive to $Re$. Once this shift is accounted for, drag reduction decreases with $Re$ at a markedly slower rate. If the slightly favorable trend of the energy required to create the forcing is considered, a chance emerges for positive net energy savings also at large values of the Reynolds number. 
\end{abstract}


\maketitle

\section{Introduction}
\label{sec:intro}

The increase of friction drag above the laminar value is one of the fundamental manifestations of turbulence in the simplest wall-bounded flows, prompting the study of techniques aimed at skin-friction drag reduction in the turbulent regime. The control strategies currently under development range from passive techniques (a classic example being riblets \cite{garcia-jimenez-2011}) that yield up 8--10\% reduction of skin friction in well controlled, low-Reynolds laboratory conditions, to reactive techniques that exploit linear control theory \cite{kim-bewley-2007} and promise much better performances, especially in terms of net energy savings, but are highly complex and so far have been studied mostly through high-fidelity Direct Numerical Simulations (DNS), where the actuator is introduced via a boundary condition and friction drag can be easily measured by a space-time averaging procedure. Between these two extrema, active predetermined techniques represent a compromise between energy expenditure and energy gain, and yield sizable net energy savings with the advantage of moderate complexity, as they do not need sensors and only require relatively large-scale actuators. One recently proposed strategy, that looks promising in terms of net energy saving potential, is the streamwise-traveling wave concept \cite{quadrio-ricco-viotti-2009}, that has been shown to provide at least 20\% net energy savings. It has been already given experimental verification \cite{auteri-etal-2010}, and studies are ongoing for promising actuator technologies \cite{gouder-potter-morrison-2013, carpi-etal-2013}. The review paper by Karniadakis and Choi \cite{karniadakis-choi-2003} focuses on spanwise-forcing techniques, and a more recent volume \cite{leschziner-choi-choi-2011} contains several papers illustrating the long-term prospects of these (and other) drag reduction techniques.

\begin{figure}
\includegraphics[width=0.8\textwidth]{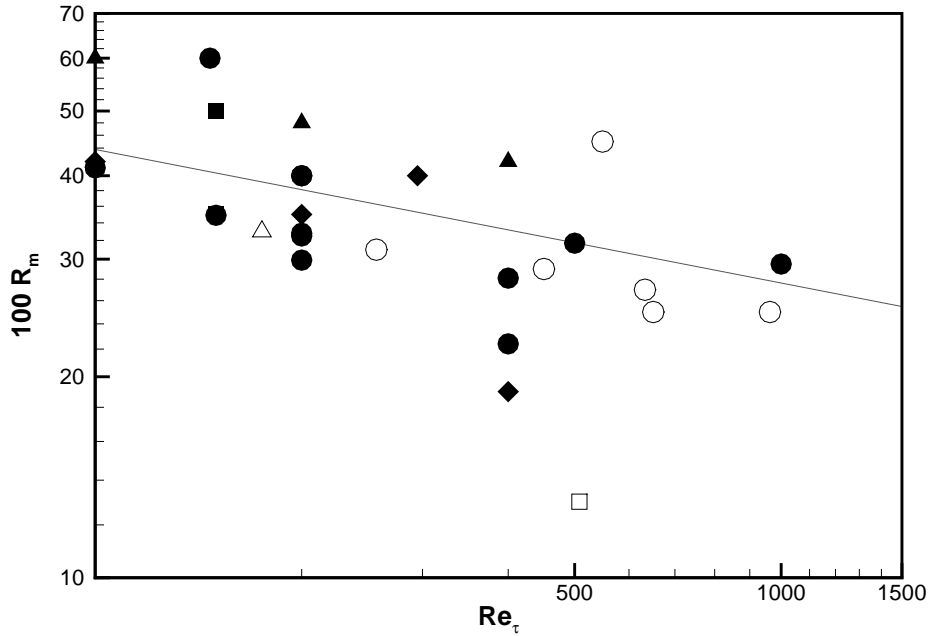}
\caption{Literature data for maximum drag reduction rate $R_m$ versus $Re_\tau$ for spanwise-forcing techniques. Black (white) symbols indicate results from DNS (experimental) studies. We explicitly note that the forcing amplitude is not always identical among different datasets. Circles: oscillating wall \cite{ricco-quadrio-2008, jung-mangiavacchi-akhavan-1992, quadrio-ricco-2004, choi-xu-sung-2002, trujillo-bogard-ball-1997, choi-graham-1998, touber-leschziner-2012, ricco-wu-2004, choi-debisschop-clayton-1998, quadrio-sibilla-2000, nikitin-2000, tamano-itoh-2012}; triangles: streamwise-traveling waves \cite{quadrio-ricco-viotti-2009, auteri-etal-2010}; squares: spanwise-traveling waves \cite{du-karniadakis-2000, du-symeonidis-karniadakis-2002}; diamonds: Lorentz force \cite{pang-choi-2004, berger-etal-2000}. The solid line is $R_m \sim Re_\tau^{-0.2}$.}
\label{fig:possiblescenarios}
\end{figure}

Riblets have already undergone flight tests \cite{viswanath-2002}, and they have been estimated to yield reductions of total aerodynamic drag of at least 2\% in flight conditions. Despite the fact that most of the anticipated applications are characterized by high values of the Reynolds number $Re$, active techniques are still lacking a comparable, thorough evaluation in high-$Re$ flows \cite{quadrio-2011}. Since the active forcing is applied at the wall, it is easy to show \cite{spalart-mclean-2011} that the percentage benefits measured at low $Re$ are expected to decrease at least as the (square root of) the baseline friction coefficient does. Indeed, the available data witness a much faster decrease of the maximum drag reduction. However, only few higher-$Re$ data exist, and they cover one $Re$ decade only, owing to the enormous increase of the computational cost of DNS, paralleled by a shrinking of the physical size and timescale of the required sensors and actuators to be employed in a laboratory experiment. 

In this work we focus on one (although rather general) class of predetermined control strategies for drag reduction: the streamwise-traveling waves of spanwise wall velocity defined by
\begin{equation}
W(x,t) = A \cos \left( \kappa_x\, x - \omega\, t \right) ,
\label{eq:wave}
\end{equation}
where $W(x,t)$ is the spanwise velocity forcing at the wall, varying with the streamwise coordinate $x$ and time $t$. The parameters of the forcing are its amplitude $A$, the wavenumber $\kappa_x$ and frequency $\omega$, which define the wavelength $\lambda=2 \pi / \kappa_x$ and the period $T = 2 \pi / \omega$. The forcing law (\ref{eq:wave}) contains the two limit cases of the spanwise-oscillating wall when $\kappa_x = 0$, and the stationary wave when $\omega =0$. Particular combinations of parameters may lead to a reduction of the skin-friction drag that we quantify, following Kasagi {\em et al.}\cite{kasagi-hasegawa-fukagata-2009}, in terms of the drag reduction rate $R$, i.e. the relative reduction in skin-friction coefficient with respect to the uncontrolled flow. How the performance of the forcing depends upon the value of the Reynolds number is typically \cite{choi-xu-sung-2002, ricco-quadrio-2008, touber-leschziner-2012,moarref-jovanovic-2012,belan-quadrio-2013} quantified in the literature through the exponent $\gamma$ of a power law $R_m \sim Re_\tau^{\gamma}$ that links the maximum drag reduction rate $R_m$, achieved at a fixed forcing amplitude $A^+$, to the value of $Re_\tau$, the Reynolds number based on the friction velocity $u_\tau$. Choi \& Graham \cite{choi-graham-1998} were first to take measurements at two different values of $Re$ in a oscillating cylindrical pipe. Their data, yielding $\gamma=-0.06$, are unfortunately of little use since the setup was enforcing a constant azimuthal displacement, thus rendering the two datasets non easily comparable owing to the variable $A^+$. Choi {\em et al.} \cite{choi-xu-sung-2002} used turbulent channel flow simulations to find that, at $T^+=100$ and $A^+=10$ (i.e. at optimal period and intermediate amplitude, as for the majority of available data), the oscillating wall produces a drag reduction of 41.1\%, 29.9\% and 22.4\% at $Re_\tau=100,200$ and $400$. They fitted several available results to infer that drag reduction is predicted by a quadratic function of a scaling parameter proportional to $Re_\tau^{-0.2}$, thus implying $\gamma=-0.4$. Ricco \& Quadrio \cite{ricco-quadrio-2008} reported that the spanwise-oscillating wall at $T^+=125$ and $A^+=12$ yields 32.5\% drag reduction at $Re_\tau=200$ and 28.1\% at $Re_\tau=400$, which corresponds to the smaller value $\gamma=-0.2$. Quadrio, Ricco \& Viotti \cite{quadrio-ricco-viotti-2009} determined the best-performing traveling wave at $Re_\tau=200$ and forcing intensity of $A^+=12$, and verified that at $Re_\tau=400$ maximum drag reduction decreases from 48\% to 42\%, thus supporting $\gamma=-0.19$. Touber \& Leschziner \cite{touber-leschziner-2012} presented DNS data at $Re_\tau=200,500$ as well as LES and one DNS datapoint at $Re_\tau=1000$ to suggest that drag reduction values obtained under similar values of a scaling parameter support $\gamma=-0.2$.

The rather large values of $\gamma$ observed by the aforementioned DNS studies imply a rapid decrease of the drag reduction effect. This message is conveyed by figure \ref{fig:possiblescenarios}, where maximum drag reduction rate $R_m$ is plotted versus $Re_\tau$ for the low-$Re$ numerical and laboratory experiments available in the literature concerning spanwise-forcing techniques. Extrapolation at higher $Re$, however, is not obvious, and alternative attempts to gather high-$Re$ information under simplifying assumptions have been reported.

The picture that emerges from the few available theoretical studies is not entirely in agreement with the empirical information. Duque-Daza et al \cite{duque-etal-2012} presented a linear stability study to link turbulent drag reduction to the growth of near-wall turbulent streaks in a laminar flow where the prescribed base flow is the mean streamwise velocity profile plus the spanwise Generalized Stokes Layer (GSL) \cite{quadrio-ricco-2011} generated by the traveling waves; although streak amplification turns out to depend on the parameters of the forcing very much like drag reduction does, small or negligible change of streak amplification is found over a wide range of $Re$. Moarref \& Jovanovi\'c \cite{moarref-jovanovic-2012} developed a model-based approach that feeds the linearized Navier--Stokes equations with DNS-computed energy statistics, and applied it to the oscillating wall. Although the $Re$-effect could be studied up to $Re_\tau=934$ only (because of the need of DNS information), they found that the maximum drag reduction decreases with $\gamma=-0.15$. Belan \& Quadrio \cite{belan-quadrio-2013} developed a perturbation analysis to predict drag reduction within an eddy-viscosity-based approach to turbulence modeling, and found $\gamma \approx 0.04$ for a bulk Reynolds number up to one million. Iwamoto et al \cite{iwamoto-etal-2005} employed both analytic developments and DNS experiments to show that a virtual drag reduction technique capable of completely removing near-wall turbulent fluctuations in a turbulent plane channel flow within the layer $y^+ < 10$ would still yield 35\% drag reduction at the relatively large value of $Re_\tau=10^5$. Although Iwamoto {et al.} suggested a logarithmic decay of drag reduction with $Re_\tau$, a power-law fits their results equally well over several decades of $Re_\tau$ and yields the value $\gamma = 0.045$. 

In this paper, we use DNS to obtain information on the higher-$Re$ performance of the spanwise-oscillating wall and the streamwise-traveling waves. One important distinguishing feature of this study is that we aim to carry out a comprehensive parametric survey at higher $Re$. In fact, we notice that most of the available studies, besides being limited to the oscillating wall, only track the neighborhood of the forcing parameters that deliver maximum drag reduction at low-$Re$, thus implicitly assuming that the optimal forcing conditions do not change with $Re$ when properly scaled in wall units. Since the computational cost may rapidly become overwhelming, our strategy to make this parametric study possible consists in carefully adjusting the size of the computational domain while monitoring the effect of this crucial discretization parameter on the reliability of the obtained drag reduction information. 

The structure of the paper is as follows. In Section \ref{sec:method} the numerical strategy and simulation parameters are described in detail, focusing on the choice of the size of the computational domain, and on the strategy adopted to quantify the error related to finite averaging time on the skin-friction coefficient. In Section \ref{sec:R} drag reduction properties for oscillating wall and traveling waves up to $Re_\tau=2100$ are discussed and the effect of $Re$ quantified. Section \ref{sec:S} addresses power consumption and power budget, and Section \ref{sec:discussion} contains a discussion of the results, followed by some conclusions in Section \ref{sec:conclusions}.

\section{Method}
\label{sec:method}

\subsection{Code and simulation parameters}

The analysis takes advantage of a newly created DNS dataset for a turbulent plane channel flow modified by either spanwise wall oscillations or streamwise-traveling waves, Eq.(\ref{eq:wave}). The DNS computer code, its parallel algorithm and the architecture of the computing system used for many of the calculations presented in this study and hosted at the University of Salerno, have been described elsewhere \cite{luchini-quadrio-2006}. The code is a mixed-discretization parallel solver of the incompressible Navier--Stokes equations, based on Fourier expansions in the homogeneous directions and high-order explicit compact finite-difference schemes in the wall-normal direction. If exception is made for the domain size, discussed below, the present set of calculations is quite standard. The governing equations are integrated forward in time, starting from an initial condition of fully developed uncontrolled channel flow, generated specifically for each Reynolds number, while the flow rate is kept constant. The Reynolds number is defined as $Re_P = U_P h / \nu$ where $h$ is half the distance between the channel walls and $U_P$ is the centerline velocity of a laminar Poiseuille flow with the same flow rate. Parametric DNS studies have been carried out at $Re_P = 4760$ and $23500$, corresponding to $Re_\tau = u_\tau h / \nu \approx 200$ and $1000$ respectively, where $u_\tau$ is the friction velocity of the non manipulated flow. A more limited dataset is produced at $Re_P = 73000$, corresponding to $Re_\tau \approx 2100$. The spatial resolution (number of Fourier modes $N_x$ and $N_z$ in the homogeneous streamwise and spanwise directions before expansion for dealiasing, and number of points $N_y$ in the wall-normal direction) is set according to current practice; the resolution improves further in flows with drag reduction. Discretization parameters for the 3 considered values of $Re$ are shown in Table \ref{tab:discretization-parameters}.

\begingroup
\squeezetable
\begin{table}
\begin{tabular*}{\textwidth}{@{\extracolsep{\fill}} c |c c c c c c c}
\hline
$Re_p$	& $L_x/h$	& $L_z/h$	& $L_x^+$ & $L_z^+$	& $N_x \times N_y \times N_z$ &  $Re_\tau$\\

4760	  &  6.28	    & 3.14	 & 1250 & 625 & $128 \times 100 \times 128$  & 199\\	
29500   &  1.32     & 0.66   & 1255 & 627 & $128 \times 500 \times 128$  & 951 \\	
73000   &  1.193     & 0.596  & 2514 & 1257 & $192 \times 1000 \times 192$ & 2108\\ 	
\hline
\end{tabular*}
\caption{Computational domain size, spatial resolution and friction Reynolds number $Re_\tau$ for the 3 sets of simulations carried out at different values of $Re_P$.}
\label{tab:discretization-parameters}
\end{table}
\endgroup

\begin{figure}
\centering
\includegraphics[width=0.8\textwidth]{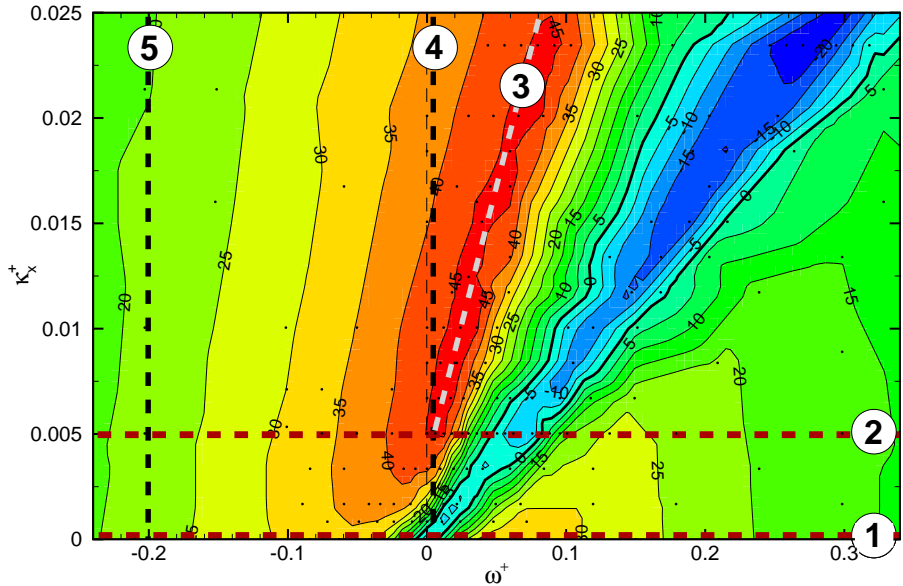}
\caption{Map of the drag reduction rate 100 $R$ versus the wavenumber $\kappa_x^+$ and frequency $\omega^+$ of the traveling waves at $A^+=12$ and $Re_\tau = 200$, after Quadrio {\em et al.} \cite{quadrio-ricco-viotti-2009}. As in the original paper, the DNS datapoints (small dots) are first linearly interpolated onto a finer regular grid that is used for contouring. The contour level is indicated on the isolines, spaced by 5. The dashed lines highlight the portions of the parameter space investigated in this work.}
\label{fig:QRV09map}
\end{figure}

The parameters defining the forcing expressed by Eq.(\ref{eq:wave}) are chosen to facilitate comparison of the results with those already available: in particular, a constant forcing amplitude at $A^+=12$ is considered throughout this study. Fig.\ref{fig:QRV09map} plots the available dataset \cite{quadrio-ricco-viotti-2009} at $Re_\tau=200$ and $A^+=12$ along with five dashed lines that mark the regions where we concentrated our analysis at $Re_\tau=200$ and $Re_\tau=1000$. Two scans of the parameter space have been made at constant wavenumber, one for the oscillating wall case at $\kappa_x^+=0$ (line 1) and one at $\kappa_x^+ = 0.005$ (line 2), where at low $Re$ the maximum drag reduction is achieved. One more scan along the locus of largest drag reductions (line 3), as well as two scans at constant frequency $\omega^+ = 0.012$ (line 4) and $\omega^+ = -0.2$ (line 5), complete the dataset. We notice that for $\kappa_x^+ = 0.005$, the wavelength $\lambda_x^+ = 1250$ of the forcing and the streamwise length of the computational domain coincide. This is of no concern, since previous studies \cite{quadrio-ricco-viotti-2009} have verified the absence of subharmonic effects.

\subsection{Performance indicators}
The control performance is evaluated, according to the notation introduced by Kasagi {\em et al.}\cite{kasagi-hasegawa-fukagata-2009}, in terms of three dimensionless indicators $(R, P_{in}, S)$. $R$ is the drag reduction rate, equivalent to the relative reduction of pumping power $P$ per unit channel area
\begin{equation}
R = \frac{P_0 - P}{P_0},
\label{eq:R}
\end{equation}
where the subscript $0$ refers to the uncontrolled flow. The time-averaged pumping power per unit channel area is computed as:
\[
P = \frac{U_b}{T_{av} \,L_x \,L_z}\int_{t_i}^{t_f} \int_{0}^{L_x} \int_{0}^{L_z} \tau_x \mathrm{d}x\,\mathrm{d}z\,\mathrm{d}t ,
\]
where $\tau_x$ is the streamwise component of the wall shear-stress, $U_b$ is the bulk velocity, held constant in the simulations, and $T_{av}=t_f - t_i$ is the interval for time average. For the present simulations where $U_b$ is constant, the drag reduction rate $R$ equals the reduction of the skin-friction coefficient $C_f$. 

The power required to create the wall forcing is computed by neglecting the mechanical losses of the actuation devices, and expressed as a fraction of the pumping power $P_0$ in the uncontrolled case:
\begin{equation}
{P}_{in} = \frac{1}{P_0 \,T_{av} \,L_x \,L_z}\int_{t_i}^{t_f} \int_{0}^{L_x} \int_{0}^{L_z} W \tau_z \mathrm{d}x\,\mathrm{d}z\,\mathrm{d}t
\label{eq:Pin}
\end{equation}
where $\tau_z$ is the spanwise component of the wall-shear stress and $W$ the imposed spanwise wall velocity. Finally, a net energy saving rate, i.e. the balance between the benefits and costs of the control, can be easily defined as $S=R-P_{in}$. The symbols $R_m$ and $S_m$ are used to denote the maximum of $R$ and over the forcing parameters, at fixed forcing amplitude. 

\subsection{Effects of the size of the computational domain}

The size (in the homogeneous directions) of the computational domains employed in the present study, shown by Table \ref{tab:discretization-parameters}, is smaller than the one usually thought to yield size-independent results, although it is several times larger than the Minimal Flow Unit \cite{jimenez-moin-1991} described by Jim\'enez and Moin as the minimal computational domain capable of sustaining the near-wall turbulence cycle. The rationale for choosing small domains is quite simple: making the domain smaller for a given small-scale spatial resolution reduces the number of unknowns in the calculation and thus makes the simulation run faster. This is only an apparent saving, though, since the averaging time required to get converged statistics correspondingly increases. Simple math \cite{jimenez-2003} shows that, when the DNS code in the wall-parallel directions uses Fourier discretization and fast Fourier transforms for computing the non-linear terms pseudo-spectrally, the computational cost for one timestep is proportional to $N_x N_z \log(N_x N_z)$. The quantity of primary interest in this work is the mean skin-friction coefficient $C_f$, defined as $C_f = 2 \aver{\tau_x} / \rho\,U_b^2$, where $\aver{ \cdot }$ is the expected value operator. Its standard deviation $\sigma_{C_f}$ is proportional to:
\begin{equation}
\sigma_{C_f} \sim \frac{\sigma_{C_f (t)}}{\sqrt{T_{av}}}
\end{equation}
where $T_{av}$ is the averaging time, $C_f(t)$ is the instantaneous space-averaged skin-friction coefficient and $\sigma_{C_f(t)}$ its standard deviation, proportional to $(L_x \times L_z)^{-1/2}$. It is thus evident that the smaller computational cost per timestep and the larger number of timesteps required to  obtain statistics of the same quality tend to compensate each other. In the present kind of simulations, however, an initial transient \cite{quadrio-ricco-2003} exists where the friction starts from the reference value of the unforced velocity field used as initial condition, and progressively reduces under the action of the drag-reducing technique. A significant amount of computing time is thus wasted for computing the initial transient after which a meaningful time averaging can be started. Using a smaller domain allows us to considerably reduce the cost of computing this initial part. 

This significant computational advantage notwithstanding, the obtained mean values may still be size-dependent. More important, since as stated above we want statistics ``of the same quality'', such quality must be somehow quantified, for example in terms of confidence interval of the mean. A strategy is thus needed to estimate the uncertainty due to the finite averaging time on the measured skin-friction and consequently on the drag reduction rate $R$.

The method employed here is based on the assumption that the time history of space-averaged friction, once the initial transient has been properly discarded, contains data which are realizations, uniformly spaced in time, of a continuous statistically-stationary (ergodic) random process. Under this hypothesis, the standard deviation $\sigma_{C_f}$ can be related to the temporal autocorrelation function $\rho(\Delta t)$ of $C_f(t)$ as follows:
\begin{equation}
\sigma^2_{C_f} = \frac{2 \sigma^2_{C_f(t)}}{T_{av}} \int_0^{T_{av}} \* \rho \left( \Delta t \right) 
\left( 1 - \frac{|\Delta t|}{T_{av}} \right) \mathrm{d} \Delta t
\label{eq:variance}
\end{equation}
where $\rho(\Delta t)= \aver{ C_f(t) C_f(t + \Delta t) } / \sigma_{C_f(t)}^2$. If $|\Delta t| \ll T_{av} $, as can be safely assumed in the present case, the above expression reduces to
\begin{equation}
\sigma^2_{C_f} = 2 \frac{\sigma_{C_f(t)}^2}{T_{av}} \mathcal{T} ,
\label{eq:variance-approx}
\end{equation}
where $\mathcal{T} =\int_0^{T_{av}} \* \rho (\Delta t) \mathrm{d}\Delta t$ is the integral timescale of the process. The variance $\sigma^2_{C_f}$ and the autocovariance involved in the definitions (\ref{eq:variance}) and (\ref{eq:variance-approx}) are themselves unknown and cannot be measured directly, hence their best estimators, i.e. the sample variance and autocovariance, are used instead. 

The standard uncertainty $s_R$ of the drag reduction rate $R$ is computed by propagating the sample standard deviations of the mean skin-friction for the uncontrolled and controlled case respectively, assuming they are independent variables. A confidence interval for $R$ can be obtained thanks to the central limit theorem, hence: 
\[
R - z_{\alpha /2} s_R \leq \aver{R} \leq R + z_{\alpha /2} s_R
\]
where $z_{\alpha /2}$ is the standardized confidence interval of a normalized Gaussian PDF and depends on the desired confidence level $1-\alpha$. In this work the commonly employed confidence level of 0.95 is chosen.

\begin{figure}
\includegraphics[width=0.8\textwidth]{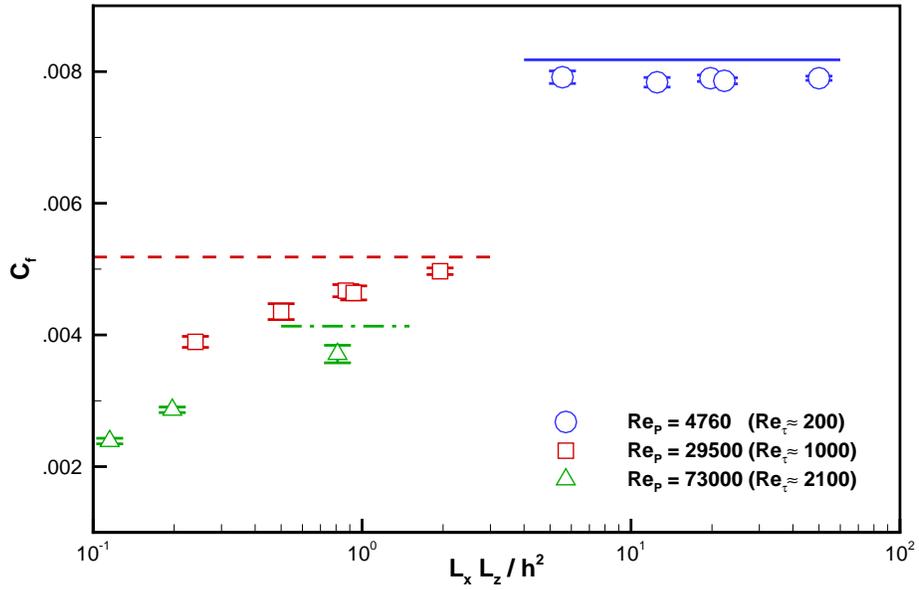}
\caption{Friction coefficient $C_f$ versus the size $L_x L_z$ of the computational domain in the homogeneous directions, expressed in outer units. Horizontal lines are the values of $C_f$ predicted by the Dean's correlation at $Re_P = 4760$ (solid blue), $Re_P = 29500$ (dashed red) and $Re_P = 73000$ (dot-dashed green). Blue circles: $Re_P=4760$; red squares: $Re_P=29500$; green triangles: $Re_P=73000$. Error bars at 95\% confidence level.}
\label{fig:Cf-area}
\end{figure}

\begin{figure}
\includegraphics[width=0.8\textwidth]{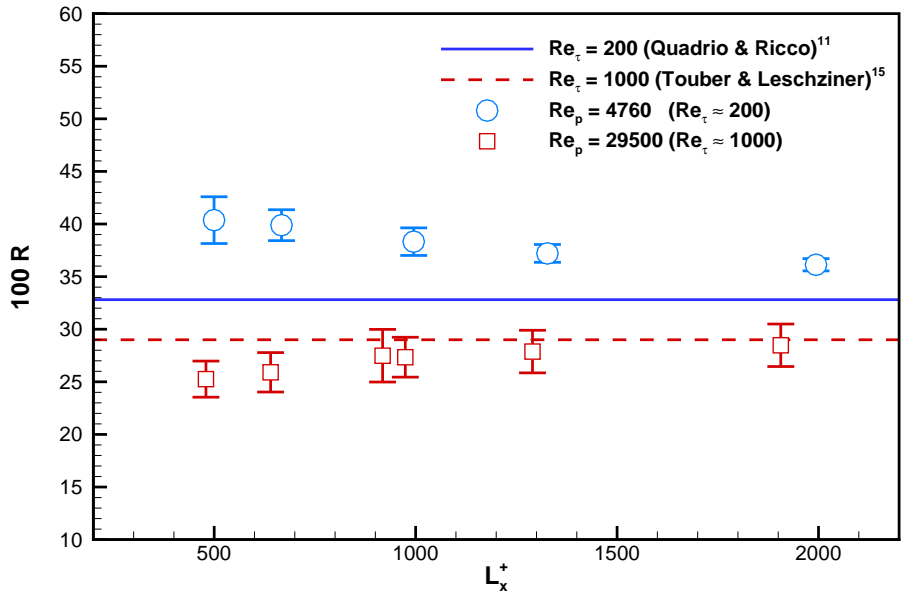}
\caption{Drag reduction rate 100$R$ versus the streamwise length $L_x^+$ of the computational domain, expressed in wall units, for the oscillating wall with $A^+=12$ and $T^+=100$. Horizontal lines are from Quadrio \& Ricco \cite{quadrio-ricco-2004} at $Re_\tau=200$ and from Touber \& Leschziner \cite{touber-leschziner-2012} at $Re_\tau=1000$. Error bars at 95\% confidence level.}
\label{fig:R-area}
\end{figure}

Figure \ref{fig:Cf-area} shows for the uncontrolled flow at the 3 considered values of $Re$ how $C_f$ depends on the domain size. This information is computed by running additional simulations with domain sizes both smaller and larger than those listed in Tab.~\ref{tab:discretization-parameters}, while the spatial resolution and aspect ratio $L_x / L_z$ of the computational domain are kept constant. Throughout the paper, line colors and symbols shapes are used to encode the value of $Re_P$: blue circles for $Re_P=4760$ ($Re_\tau \approx 200$), red squares for $Re_P=27000$ ($Re_\tau \approx 1000$) and green triangles for $Re_P=73000$ ($Re_\tau \approx 2100$). 

It is seen that the computed values of $C_f$ tend to the prediction by Dean's correlation \cite{dean-1978} as the domain size increases, and that too small domain sizes yield underestimated values of $C_f$. This is consistent with previous information \cite{jimenez-moin-1991} as well as with the observation \cite{flores-jimenez-2010} that turbulent fluctuations are progressively damped in smaller computational domains. At the lowest value of $Re_\tau=200$, there is no apparent change of $C_f$ with the considered domain sizes, which are always relatively large in outer units. The small difference between the numerical prediction and the Dean's-computed value can be attributed to the known slight inaccuracy of the latter at low $Re$: for example Kim {\em et al.} \cite{kim-moin-moser-1987} found the Dean's value to be higher by approximately 3\%. 

Although the absolute values of $C_f$ are certainly important, even more important in this study is the correct prediction of the drag reduction rate $R$. Owing to cancellation of the systematic bias related to the domain size, $R$ might be less affected by the domain size. Figure \ref{fig:R-area} shows the effect of gradually reducing the domain size on the computed drag reduction achieved by wall oscillations with $A^+=12$ and $T^+=100$. 
At both values of $Re$, increasing the domain size produces $R$ that approaches the full-domain value. The relatively large error in $R$ for the oscillating wall at $Re_\tau=200$ near the optimal oscillating conditions considered in figure \ref{fig:Cf-area} should be regarded as a worst-case: either far from the optimum, or by considering the traveling waves instead of the oscillating wall leads to significantly better estimates. 

\section{Drag reduction}
\label{sec:R}

This Section presents the results obtained for the drag reduction rate $R$, by comparing our DNS results with available literature information. In the figures, present data are shown with empty symbols, whereas literature DNS data obtained with domains of regular size are shown with filled symbols. Error bars are computed at $95\%$ confidence level; they are not visible when their size becomes smaller than the symbol. 

\subsection{Oscillating wall}

\begin{figure}
\centering
\includegraphics[width=0.8\textwidth]{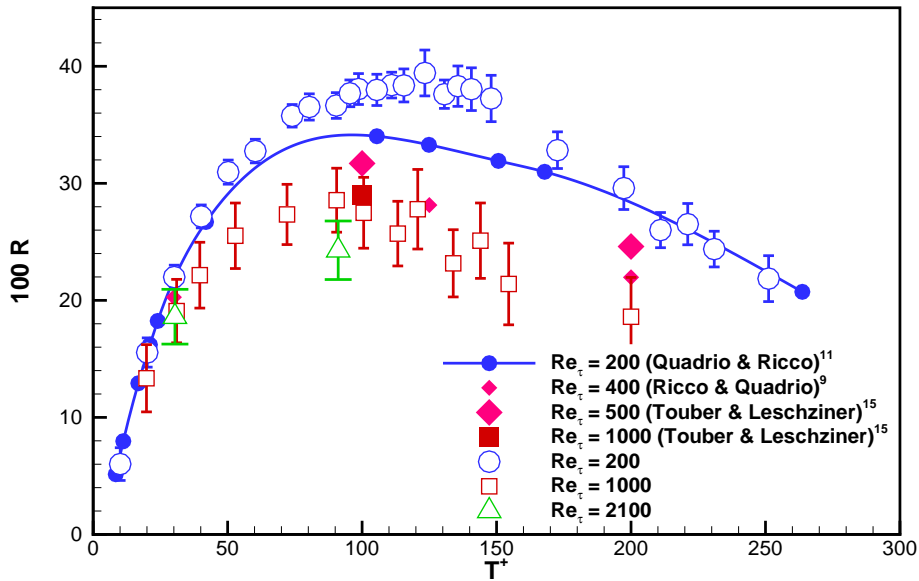}
\caption{Percentage drag reduction rate $100 R$ versus oscillation period $T^+$ for the oscillating wall at $A^+=12$ (line 1 in figure \ref{fig:QRV09map}). Open symbols: present DNS data at $Re_\tau=200, 1000$ and $2100$. Filled symbols: available literature data \cite{quadrio-ricco-2004, ricco-quadrio-2008, touber-leschziner-2012}. Error bars at $95\%$ confidence level.} 
\label{fig:R-oscillatingwall}
\end{figure}

Figure \ref{fig:R-oscillatingwall} plots $100 R$ along the horizontal line numbered (1) in figure \ref{fig:QRV09map}, i.e. drag reduction rate as a function of the period $T^+$ of the oscillating wall at $A^+=12$. Additional literature data at different $Re$ are plotted as specified in the legend. In particular, the low-$Re$ DNS dataset at $Re_\tau=200$ by Quadrio and Ricco \cite{quadrio-ricco-2004} is indicated with line-connected circles. Additional datapoints are those at $Re_\tau=500$ and $Re_\tau=1000$ by Touber and Leschziner \cite{touber-leschziner-2012}, and that at $Re_\tau=400$ from Ricco and Quadrio \cite{ricco-quadrio-2008}. Only at the lowest $Re$ the literature DNS data span a significant part of the parameter space, whereas at higher $Re$ the parameter space is sampled much more sparsely.

The values of $R$ obtained in the present study with simulations at $Re_\tau=200$ agree quite well with those from the full-channel simulations \cite{quadrio-ricco-2004} at the same $Re$. The agreement is almost perfect at values of $T^+$ much larger and much smaller than the optimum value. Near the optimum, the position of the maximum is well captured, but a slight overestimation of the drag reduction is observed,  as already shown in figure \ref{fig:R-area}, together with an increase of the uncertainty.

At $Re_\tau = 1000$ our dataset shows that $R_m$ drops from 0.39 to 0.29, and occurs at $T^+ = 90$. In spite of the non-negligible uncertainty level, we notice that the optimal parameters of the wall forcing slightly shift towards shorter periods. The only available DNS point at this $Re$ is Touber \& Leschziner's one and is quite in agreement with our dataset. 
The changes in drag reduction for increasing $Re$ seems to become smaller at smaller $T^+$. For example at $T^+=30$ $R$ decreases from $0.22$ at $Re_\tau=200$ to $0.19$ at $Re_\tau=1000$. The DNS from Ricco \& Quadrio \cite{ricco-quadrio-2008} carried out at $Re_\tau=400$ and the same value of $T^+$ supports this observation.

This general picture is confirmed by the observation of the two datapoints available at $Re_\tau=2100$, which show very small decrease of performance at small $T^+$ and more substantial one near the optimal oscillation period, where $R_m$ becomes 0.24.

\subsection{Streamwise-traveling waves}

\begin{figure}
\includegraphics[width=0.8\textwidth]{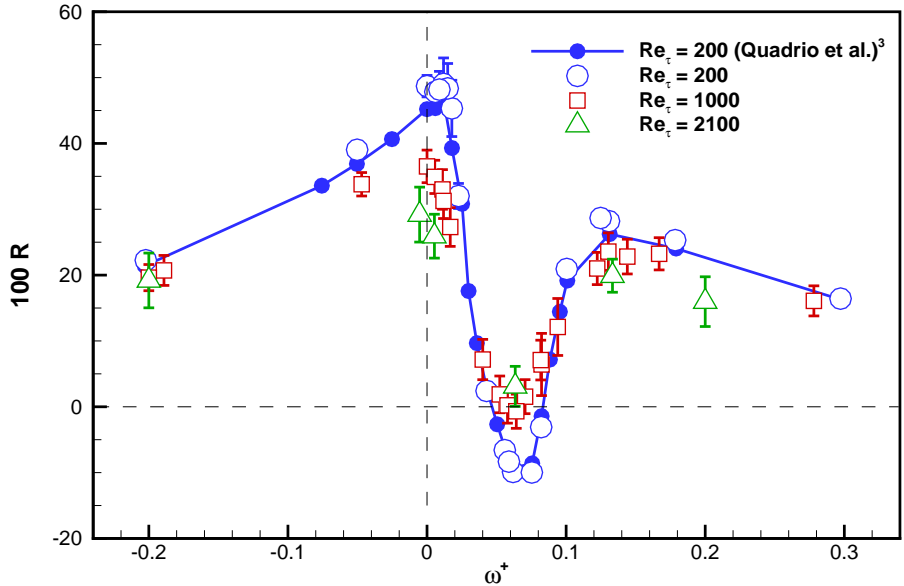}
\caption{Percentage drag reduction rate $100 R$ versus oscillating frequency $\omega^+$ for the streamwise-traveling waves at $A^+=12$ and $\lambda^+=1250$ (line 2 of figure \ref{fig:QRV09map}). Symbols and error bars as in figure \ref{fig:R-oscillatingwall}.}
\label{fig:R-travelingwave-constk}
\end{figure}

Figure \ref{fig:R-travelingwave-constk} plots $100 R$ along the horizontal line numbered (2) in figure \ref{fig:QRV09map}, i.e. drag reduction as a function of the forcing frequency $\omega^+$ for traveling waves with constant wavelength $\lambda^+=1250$ and $A^+=12$. The line passes near the known \cite{quadrio-ricco-viotti-2009} maximum drag reduction $R_m=0.48$ at $Re_\tau=200$; here at the same $Re$ we measure $R_m=0.49$, the small difference being clearly within the confidence interval. Overall, there is nearly perfect agreement between the presently computed data at $Re_\tau=200$ and the available results: the overestimate of drag reduction when the computational domain is not fully adequate is much weaker here than for the oscillating wall. Moreover, as in that case, the effect is confined to the region near maximum drag reduction. 

As the Reynolds number is increased to $Re_\tau=1000$, $R_m$ drops to 0.37, with a total loss of 0.12. A similar change of $R$, although of opposite sign, can be observed in the drag-increasing ``valley''. The drag increase almost disappears at higher $Re$, but a local minimum of $R$ is still present and the concave part of the curve seems to widen, embracing a larger range of (positive) $\omega^+$. At this wavelength, the best wave seems to be the stationary wave at $\omega^+=0$, at odds with lower $Re$, where the maximum drag reduction is achieved for a small positive frequency. At large positive and negative frequencies, and in particular for $|\omega^+| > 0.15$, $R$ appears to be almost unchanged by the increased $Re$. The more limited dataset at $Re_\tau=2100$ supports the trends discussed above. The highest drag reduction decreases further down to 0.29. No drag increase is observed in the ``valley'' where a small drag reduction rate of approx. 0.03 is achieved instead.

\begin{figure}
\includegraphics[width=0.8\textwidth]{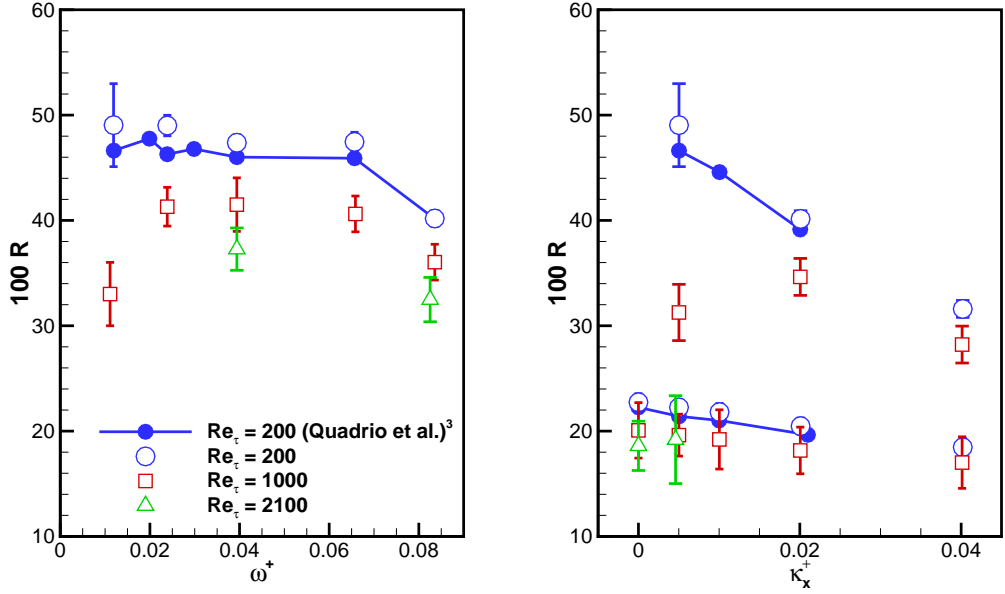}
\caption{Percentage drag reduction rate $100 R$ for the streamwise-traveling waves at $A^+=12$ along lines (3), (4) and (5) of figure \ref{fig:QRV09map}. Left: line (3), maximum drag reduction ridge. Right: lines (4) at $\omega^+=0.012$ (top points) and line (5) at $\omega^+=-0.20$ (bottom points). Symbols and error bars as in figure \ref{fig:R-oscillatingwall}.}
\label{fig:R-travelingwave-constomega}
\end{figure}

In figure \ref{fig:R-travelingwave-constomega}, on the left, the interest is focused on the low-$Re$ ridge of maximum drag reduction, indicated with line (3) in figure \ref{fig:QRV09map}. Again, at low $Re$ our data confirm the available DNS information, with only a very slight overestimate (and a large error bar) of $R$ near the maximum. The low-frequency low-wavenumber region of the ridge is strongly affected by an increase in $Re$, whereas the higher-frequency part is much less sensitive to it. The maximum $R$ at $Re_\tau = 1000$ is 0.42 and significantly shifts towards higher frequencies.  

Lastly, figure \ref{fig:R-travelingwave-constomega} on the right reports the data available along the constant-frequency lines (4) and (5) of figure \ref{fig:QRV09map}. Line (4) is drawn in a region of large drag reduction, corresponds to a very small positive frequency, and passes through the low-$Re$ point corresponding to $R_m$. The data confirm that $R$ is strongly affected by $Re$ in the neighborhood of the low-$Re$ maximum. At the same time, in the region of higher wavenumbers the effect of increasing $Re$ becomes rather mild, and the same happens along line (5), which is quite far from the low-$Re$ optimum. All our datapoints there show a very modest reduction of performance.

\section{Power budget}
\label{sec:S}

\begin{figure}
\centering
\includegraphics[width=0.8\textwidth]{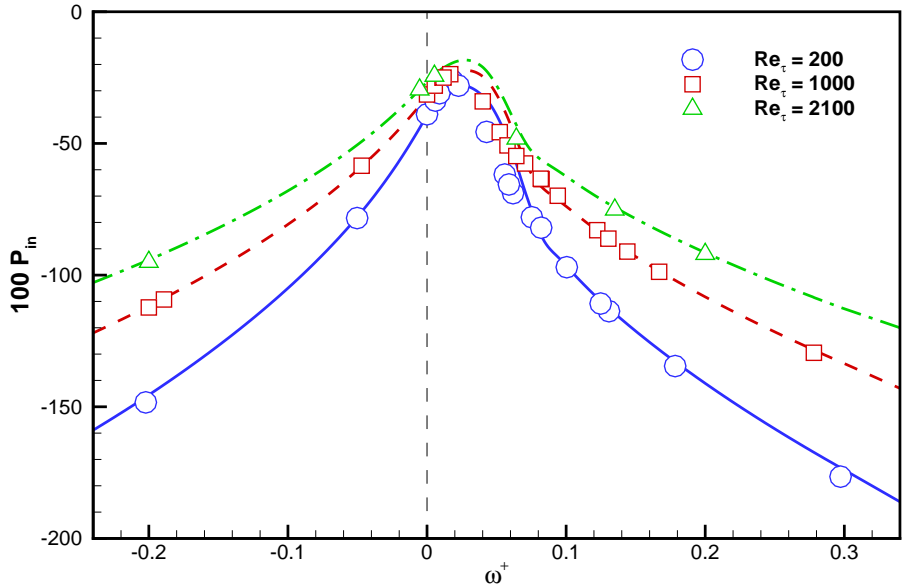}
\caption{Input power $P_{in}$ normalized with pumping power $P_0$ versus frequency $\omega^+$ for the streamwise-traveling waves at $A^+=12$ and $\lambda^+=1250$. Symbols as in figure \ref{fig:R-oscillatingwall}. Lines are computed from the laminar analytic solution at $Re_\tau=200$ (solid blue), $Re_\tau=1000$ (dashed red) and $Re_\tau=2100$ (dot-dashed green).}
\label{fig:Pin-travelingwave-constk}
\end{figure}

Figure \ref{fig:Pin-travelingwave-constk} illustrates the nondimensional power input $P_{in}$, defined by Eq.(\ref{eq:Pin}) as the power required by the wall forcing, normalized with the pumping power. Only one case for traveling waves at $\kappa_x^+=0.005$ (i.e. line (2) in figure \ref{fig:QRV09map}) is shown. Results from the present DNS, indicated by symbols, are compared with the prediction based on the analytical expression of the laminar Generalized Stokes Layer (GSL) derived by Quadrio \& Ricco \cite{quadrio-ricco-2011}. In fact, the wall value of the GSL velocity profile and its wall-normal derivative completely determine $P_{in}$, provided the GSL correctly describes the mean spanwise flow in the turbulent case. By a simple manipulation of the GSL equation one writes:
\begin{equation}
P_{in} = \frac{(A^+)^2}{2 U_b^+}\mathcal{R}\left\{  Ce^{\pi \mathrm{i} /6} \left( \kappa_x^+ \right)^{1/3}  \right\}
\mathrm{Ai}^{'}\left[ -e^{\pi \mathrm{i} /6}\left( \kappa_x^+ \right)^{1/3} \left( \frac{\omega^+}{\kappa_x^+} + \mathrm{i} \kappa_x^+  \right)\right]
\label{eq:Pin-GSL}
\end{equation}
where $\mathcal{R}$ indicates the real part, $\mathrm{Ai}^{'}$ is the first derivative of the Airy function of the first kind, $\mathrm{i}$ is the imaginary unit and $C=\left\{ \mathrm{Ai}\left[ \mathrm{i}e^{i\pi/3}(\kappa_x^+)^{1/3} (\omega^+/\kappa_x^+ + \mathrm{i}\kappa_x) \right] \right\}^{-1}$ is a constant.  

The rather good agreement between DNS data and Eq.(\ref{eq:Pin-GSL}) confirms that the GSL in the turbulent flow does not differ significantly from the laminar solution even at higher values of $Re$, although regions of the parameter space do exist where the laminar prediction is not perfect. This happens, as expected \cite{quadrio-ricco-2011}, in the drag-increasing region where the waves travel forward with a phase speed comparable to the convective speed of the near-wall velocity fluctuations \cite{quadrio-luchini-2003}.

Figure \ref{fig:Pin-travelingwave-constk} highlights an important characteristic of the laminar GSL: $P_{in}$ generally decreases with $Re$. As shown in Eq.(\ref{eq:Pin-GSL}), $P_{in}$ varies with $Re_\tau$ only because of a change of $U_b ^+$, which can be expressed, following Pope \cite{pope-2000}, as $U_b^+ = 7.715\, Re_{\tau}^{0.136}$. This quantitatively confirms and extends to the traveling waves the suggestion that $P_{in}$ should decrease as $P_{in} \sim Re_\tau^{-0.136}$, already put forward for the oscillating wall \cite{ricco-quadrio-2008}. This brings about interesting perspectives for the net energy saving rate $S=R-P_{in}$: indeed, in several parts of the control parameter space $R$ decreases slower than $P_{in}$, implying that $S$ actually increases with $Re$. 

\begin{figure}
\centering
\includegraphics[width=0.8\textwidth, trim = 0mm 10mm 0mm 0mm]{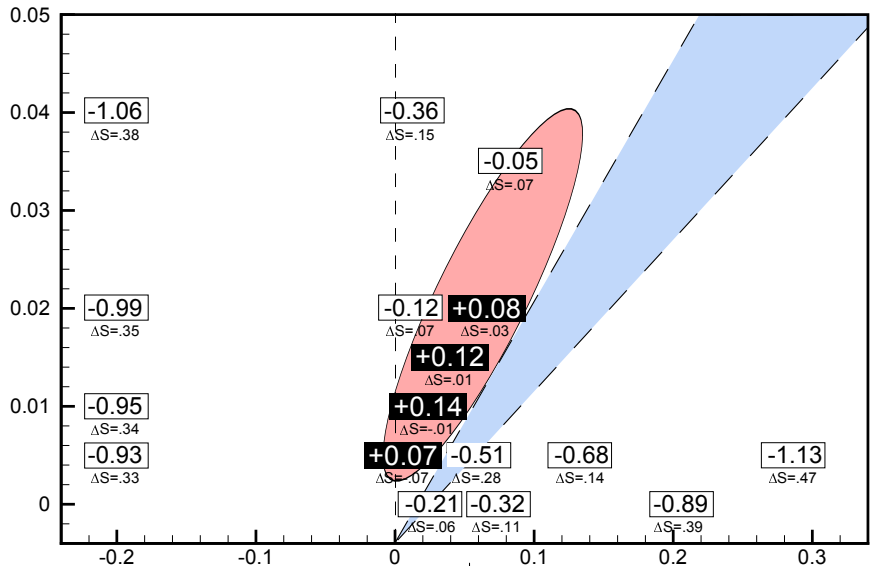}
\caption{Net energy saving rate $S$ at $Re_\tau=1000$ for the traveling waves with amplitude $A^+=12$. Compared to figure \ref{fig:QRV09map}, the vertical axis is extended to include larger wavenumbers. The shaded areas mark two important low-$Re$ regions: the light-red ellipse encloses the region of large drag reduction, while the light-blue triangle marks the region of drag increase. Rectangle-enclosed figures quantify the local value of $S$, with filled rectangles highlighting positive $S$. Below them, the smaller numbers quantify $\Delta S$, the change of $S$ when $Re_\tau$ increases from 200 to 1000.}
\label{fig:S-Re1000}
\end{figure}

Figure \ref{fig:S-Re1000} plots $S$ for traveling waves at $Re_\tau=1000$ and $A^+=12$. In the map, the positive values of $S$ are graphically emphasized by filled rectangles. $S>0$ occurs within the (light red) elliptic region delimited by the solid line, which identifies the low-$Re$ ridge of largest drag reduction. In the rest of the parameter space $S$ is negative (owing to the large value of $A$). One first notices that the positions of maximum drag reduction rate
$R_m$, minimum $P_{in}$ and maximum energy saving $S_m$ are not the same, whereas at  $Re_\tau=200$ these points coincide at $\kappa_x^+= 0.005$ and $\omega^+=0.018$. At  $Re_\tau=1000$, $S_m$ moves to $\kappa_x^+= 0.01$ and $\omega^+=0.024$, and $R_m$ moves to $\kappa_x^+= 0.014$ and $\omega^+=0.04$, while the position of minimum $P_{in}$ is unchanged owing to its perfect wall units scaling. Hence, the shift of $S_m$ is only due to the large $Re$-sensitivity of $R$ in the low-$\omega^+$, low-$\kappa_x^+$ part of the drag reduction ridge, which causes $R_m$ to shift towards higher frequencies and wavenumbers. The scenario is fully confirmed by the limited dataset at $Re_\tau = 2100$ (not reported in the figure), for which $S_m=0.10 \pm 0.02$ occurs at $\kappa_x^+= 0.015$ and $\omega^+=0.04$.

The positive trend of $S$ with $Re$, discussed before in the context of figure \ref{fig:Pin-travelingwave-constk}, is demonstrated in figure \ref{fig:S-Re1000} by the numerical value of $\Delta S$, i.e. the change in $S$ when $Re_\tau$ is increased from 200 to 1000. $\Delta S$ is positive for every data point at $S<0$, usually at high $\omega^+$ or distant from the drag reduction ridge. 

We would like to underline that the numerical values of $S$ should not be emphasized too much. Indeed, the entire dataset is computed for the sole value of $A^+=12$, whereas it is well known that the best $S$ are obtained at smaller amplitudes. Moreover, $S$ is defined as a difference between $R$ and $P_{in}$: while $P_{in}$ does not present perceivable statistical uncertainty, we have seen that $R$ does. Hence the numerical values of $S$, whose magnitude can be smaller than $R$, suffer from a larger relative uncertainty, so that the discussion above is only intended to highlight the interesting positive trend of $S$ with $Re$.

\section{Discussion}
\label{sec:discussion}

The results presented so far do not disagree with the information available in the literature: when the value of $Re$ is increased, the maximum drag reduction decreases rapidly. However, the present study highlights that this is true only for the region of the parameter space where the low-$Re$ optimum is located. Indeed, the decrease rate of drag reduction significantly depends upon the particular region of parameter space, with the low-$Re$ optimum and drag increase ``valley'' showing the fastest decrease. This is a new observation, although in the available literature a few consistent datapoints can indeed be found. For example, Ricco and Quadrio \cite{ricco-quadrio-2008} in their study of the oscillating wall presented a DNS point at small $T^+$ where $R$ decreases very slightly from $Re_\tau=200$ to $Re_\tau=400$. However, a clear and general picture is emerging for the first time from the present work, where at the same time the more revealing streamwise-traveling waves are considered, and a parametric survey is carried out with a five-fold $Re$ separation and, with a more limited set of datapoints, up to ten-fold $Re$ separation.

To describe quantitatively how a change in $Re$ affects the drag reduction rate $R$, it can be assumed that, at least within the present range of Reynolds numbers, $R$ depends on $Re_\tau$ according to a simple power law:
\begin{equation}
R \sim Re_\tau^\gamma
\label{eq:simple-exponent}
\end{equation}

Such a choice has been employed in the past, hence it is a required step to compare our findings with available information; however, it is in itself a rather arbitrary choice, and alternatives could be considered. For example  Belan \& Quadrio \cite{belan-quadrio-2013} found that their predictions of $R_m$, computed for a bulk Reynolds number up to one million, are better fitted by a law of the type $R = \alpha + Re_\tau^\beta$. Obviously, both the coefficients $\alpha$ and $\beta$ could eventually be considered as functions of the parameters themselves, and/or of $Re$. However, if we stick to the functional dependence (\ref{eq:simple-exponent}), the present study clearly highlights that $\gamma$, which plays the role of a sensitivity coefficient, is not constant when $\omega$ and $\kappa_x$ are varied.

The case of the oscillating wall ($\kappa_x=0$) is somewhat simple, and not entirely revealing. The normally accepted value for the empirical coefficient, i.e. $\gamma \approx -0.20$, is confirmed by our data, but only as far as $R_m$ is concerned. On the other hand, at short periods of oscillation, like for example $\omega^+=0.2$ or $T^+ \approx 30$, $R$ is less sensitive to a change in $Re$, leading to $\gamma \approx -0.08$. 

\begin{figure}
\centering
\includegraphics[clip=true, trim=0mm 0mm -5mm 0mm,width=\textwidth]{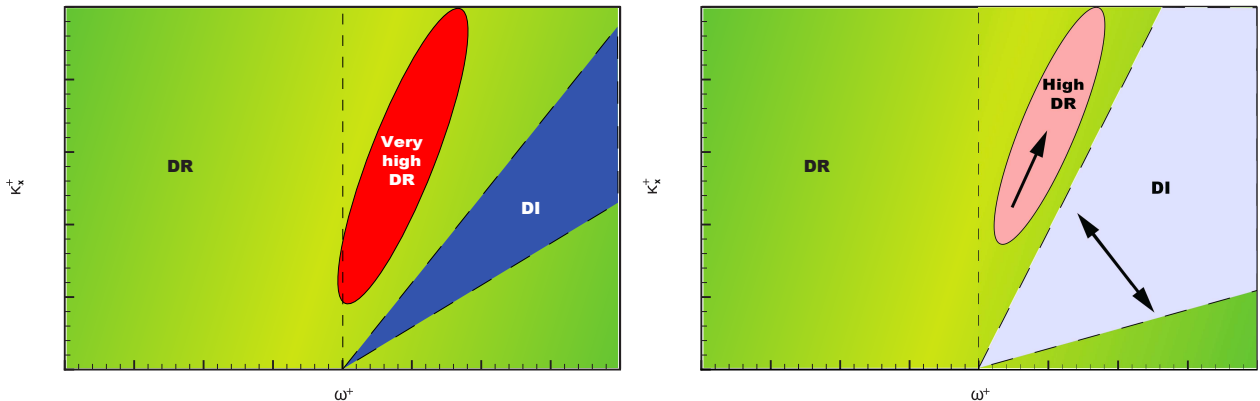}%
\caption{Sketch of the changes in drag reduction by streamwise-traveling waves in the $\kappa_x^+ - \omega^+$ plane when $Re$ increases: low-$Re$ is pictured at the left, and higher-$Re$ at the right. The color scale is similar to that of figure \ref{fig:QRV09map}, with maximum drag reduction in red and drag increase in blue. The arrows highlight the shift of the large-drag-reduction area towards higher frequencies and wavenumbers, and the widening of the drag-increase region. The remaining regions of the plane are much less affected.}
\label{fig:guess}
\end{figure}

The traveling waves present a more complex behavior, tentatively sketched in figure \ref{fig:guess}. The size of the ridge with the largest observed drag reductions shrinks when $Re$ increases, and its low-frequency, low-wavenumber tail rapidly vanishes. Here is where the measured local sensitivity is highest, with $\gamma \approx -0.25$. This part of the drag reduction map seems thus to be an essentially low-$Re$ feature, bound to disappear at large $Re$. On the other hand, the high-frequency, high-wavenumber part of the ridge is much less sensitive to $Re$, and presents $\gamma=-0.1$ (at $\omega^+=0.04$, $\kappa_x^+=0.015$) which reduces to $\gamma=-0.09$ (at $\omega^+=0.08$, $\kappa_x^+=0.035$). In this part of the ridge large values of $R$ can still be attained, ranging between 0.4 and 0.5 at $A^+=12$. This suggests the possibility that interestingly large $R$ and $S$ can still be obtained at high $Re$. Interesting is also how the low-$Re$ drag-increasing region is modified: the drag increase weakens, while the interested area appears to widen. At $Re_\tau=1000$ the drag increase at $\kappa_x^+=0.005$ almost disappears and at $Re_\tau=2100$ turns into mild drag reduction. Lastly, at high frequencies ($|\omega^+|>0.2$) both forward- and backward-traveling waves are but weakly affected by a change of $Re$: we observe $\gamma =-0.08$ at low wavenumbers and $\gamma=-0.05$ at the relatively high $\kappa_x^+=0.04$. 

\begin{figure}
\centering
\includegraphics[width=0.9\textwidth]{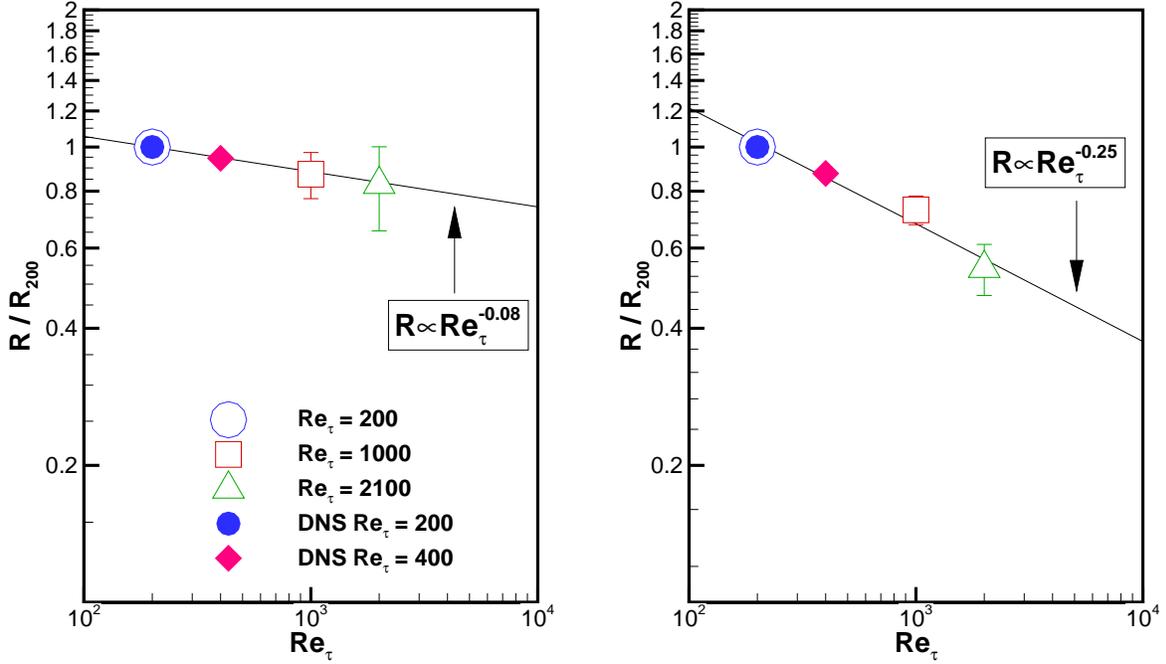}%
\caption{Relative change of $R$ as function of $Re_\tau$ for traveling waves with wavenumber $\kappa_x^+=0.005$ and amplitude $A^+=12$, at $\omega^+=-0.2$ (left) and $\omega^+=0.005$ (right). The results of two large-domain simulations at $Re_\tau=200$ and $Re_\tau=400$ are shown with filled symbols, and fully confirm the observed differences in the decrease rate of drag reduction. The region far from the low-$Re$ optimum (left) presents much lesser $Re$ sensitivity.}
\label{fig:re-effectcheck}
\end{figure}

Two weak points admittedly exist in our reasoning. The first is that we cannot be sure as to whether the observed behavior is general or concerns spanwise forcing only. Although there is no specific reason to favor the last possibility, further studies are definitely required to clarify this issue. The second one is even more important, and hinges upon the size of the spatial domain employed for the simulations. Using relatively small domain sizes has been an enabling step to make such a large parametric study possible, but the obtained results may still be dependent on the domain size, although we have made an effort to ensure that this effect is kept reasonably small and under control. To address this crucial issue at least partially, two distinct traveling waves are studied in comparative form at $Re_\tau=200$ and $Re_\tau=400$ through large, full-size DNS simulations. Both waves have $A^+=12$ and $\kappa_x^+=0.005$ but a different oscillation frequency, and thus belong to different regions of the parameter space: their drag reduction is thus expected to decrease very differently. The domain size for these simulations is chosen to be $L_x=6 \pi h$ and $L_z=3 \pi h$, as done in previous work at $Re_\tau=200$ \cite{quadrio-ricco-viotti-2009}, and the same spatial resolution in wall units is employed. When increasing $Re$, the size of the domain is kept constant in outer units and the spatial resolution is kept constant in inner units. These new data points are shown in figure \ref{fig:re-effectcheck} with filled symbols. The left plot corresponds to a case at higher (in absolute value) forcing frequency, $\omega^+=-0.2$, and lies in a region where the sensitivity is low at $\gamma=-0.08$. The right plot, on the other hand, corresponds to a case near the low-$Re$ optimum at  $\omega^+=0.005$, and in that region sensitivity attains its maximum value at $\gamma=-0.25$. It can be clearly appreciated that, in both cases, the full-size simulations yield results that entirely confirm the picture described above and in particular the existence of different regions in the parameter space where drag reduction decreases at very different rates. Hence, although the specific values of $R$ and $S$ may be slightly miscalculated within the present approach, the rate at which such quantities change for increasing $Re$ appears to be robustly computed.

Additional evidence exists that points to the substantial correctness of our description. For example, at the 9th EFMC in Rome, where we presented a preliminary version of the present work, we became aware of a closely related study (Hurst and Chung, private communication), in which the standard computational procedure with large computational domains and highly demanding numerical simulations is employed. As a consequence, the parameter space is not sampled in full detail. On the other hand, the results do not suffer from domain size effects and support our findings, confirming a strong dependence of the performance of the flow control technique on the forcing parameters. Moreover, although limited to the oscillating wall, our predictions agree with the simulation of Touber \& Leschziner \cite{touber-leschziner-2012} at $Re_\tau=1000$ in predicting the decay rate of $R_m$.

The consequences of the previously described scenario are noticeable. Some are obvious. For example, since the optimal set of control parameters is found to change with $Re$, and new regions of the parameter space may yield interesting values of drag reduction and net power saving at high $Re$, the outcome of the present study reinforces the need for experimental measurements at application-level $Re$. Since we believe it is inconceivable to increase $Re$ in DNS calculations by additional orders of magnitudes, the implication is that a suitably miniaturized actuator for a true field test is required.

One less-obvious remark descends from the observation that the results obtained in the present work by using a rather small (as measured in outer units) computational domain are well in line with those from large-scale simulations as far as changes with $Re$ are concerned. The fact poses new questions on the role of the largest turbulent structures, progressively misreprented by smaller domains, which reside away from the wall and have been proven to modulate the inner flow \cite{ganapathisubramani-etal-2012}, on the mechanisms that modulates the drag reduction with $Re$ in this range of Reynolds numbers. Moreover, we observe that the small values of $\gamma$ found here in regions far from the low-$Re$ optimum are similar to the predictions obtained from simplified linear models like the one by Duque-Daza et al \cite{duque-etal-2012}. This once again emphasizes the dynamical importance of linear processes in the near-wall region of turbulent flows.

\section{Conclusion}
\label{sec:conclusions}

This work has investigated via DNS how increasing the value of the Reynolds number from $Re_\tau=200$ up to $Re_\tau = 1000$ changes the drag-reducing properties of the streamwise-traveling waves of spanwise wall velocity. A more limited dataset at $Re_\tau=2100$ has also been presented and discussed. We considered, at each value of $Re$, several wavenumbers and frequencies; adjusting the domain size in the two homogeneous directions has been an enabling approach to limit the computational effort and to succeed in running such parametric surveys. The collateral effects of this adjustment needed to be properly addressed. On one hand, the finite-averaging-time statistical uncertainty for wall-shear fluctuations of larger amplitude has been quantified. On the other hand, a comparison of few points to DNS results obtained with large computational domains has shown that the rate of change in drag reduction with $Re$ is correctly predicted, although the specific values of the drag reduction rate $R$ might be slightly overestimated (especially when $R$ is large).

The global qualitative picture that emerges from our study is that drag reduction always decreases when $Re$ is increased, but the rate at which $R$ drops markedly depends on the control parameters $\kappa_x^+$ and $\omega^+$. The steepest decay is observed in regions of the parameter space where, at low $Re$, maximum drag reduction and drag increase occur. However, the decay is much slower in other regions, so that the optimal control parameters are shifted towards higher frequencies and wavenumbers. The control parameters yielding the minimum (nondimensional) input power $P_{in}$, maximum $R$ and maximum net power saving rate $S$ do not coincide any more at higher $Re$. For a given forcing amplitude, increasing $Re$ results in a reduction of $P_{in}$, and regions exist where the required power decays faster than $R$, thus resulting in an increase of the net power saving.

\section{acknowledgments}
We wish to thank S. Chernyshenko, Y. Chung, B. Frohnapfel and Y. Hasegawa for interesting discussions on the subject. We are particularly indebted to P. Luchini for several comments, and for the continued use of his computer system. DG is also grateful to the Fritz und Margot Faudi foundation for supporting his research period at TU Darmstadt. 

\bibliographystyle{unsrtnat}

\end{document}